\documentclass[apm,reprint,amsmath,amssymb]{revtex4-1}

\usepackage{graphicx}
\usepackage{bm}

\begin{document}

\title{Electronic structure and weak
itinerant magnetism in metallic Y$_2$Ni$_7$}

\author{David J. Singh}

\affiliation{
Department of Physics and Astronomy, University of Missori,
Columbia, MO 65211-7010, USA}

\email{singhdj@missouri.edu}


\date{\today}

\begin{abstract}
We report a density functional study of the electronic structure and
magnetism of Y$_2$Ni$_7$. The
results show itinerant magnetism very similar to that
in the weak itinerant ferromagnet Ni$_3$Al.
The electropositive Y atoms in Y$_2$Ni$_7$ serves to donate charge to the Ni
host mostly in the form of $s$ electrons.
The non-spin-polarized state shows a high density of states at the Fermi
level, $N(E_F)$ due to flat bands. This leads to the ferromagnetic
instability. However, there are also several
much more dispersive bands crossing $E(F)$, which should promote the
conductivity. 
Spin fluctuation effects appear to be comparable to or
weaker than Ni$_3$Al, based on comparison with experimental data.
Y$_2$Ni$_7$ provides a uniaxial analogue to cubic Ni$_3$Al for 
studying weak itinerant ferromagnetism, suggesting detailed measurements
of its low temperature physical properties and spin fluctuations, as well
experiments under pressure.
\end{abstract}


\maketitle

\section{Introduction}

Weak itinerant ferromagnetism is a topic of ongoing interest, both from
the point of view of understanding the physical behavior of metals
near quantum critical points, and because of the fact that these
materials often have relatively high ordering temperatures when
scaled to the ordered moment.
There is also renewed interest in itinerant magnetism because of the
unusual magnetic properties of the Fe-based superconductors,
\cite{johnston,lumsden,mazin-mag,bondino}
and spin-fluctuation pairing models for these and other unconventional
superconductors.
\cite{scalapino,mathur,johannes,moriya,mazin-spm,kuroki-spm,scalapino2}

Elemental fcc Ni metal is a classic example of an itinerant ferromagnet.
Substitution by 25\% with the trivalent element Al to form Ni$_3$Al
strongly reduces the Curie temperature to yield a weak itinerant
ferromagnet near a critical point that can be reached under pressure.
\cite{buis,niklowitz}
Both Ni$_3$Al and Ni$_3$Ga show evidence for strong exchange enhancement
and spin-fluctuation effects, including quantum spin fluctuation induced
suppression of ferromagnetism 
\cite{shimizu,moriya-book} in Ni$_3$Ga.
\cite{boer,bernhoeft,bernhoeft2,hayden,winter,niklowitz,aguayo,smith}
These two compounds have very similar electronic structures,
and differ mainly in the strength of the quantum spin fluctuations.
This difference places them on opposite sides of a ferromagnetic
quantum critical point at ambient pressure. Unusual physical
behavior including non-Fermi liquid scalings extending to very low
temperature has been found in Ni$_3$Al under pressure. \cite{niklowitz}

Y$_2$Ni$_7$ forms in a rhombohedral (R$\bar{3}$m) Gd$_2$Co$_7$ structure
\cite{virkar,buschow1,colinet}
and is a ferromagnet.
\cite{buschow,lemaire,inoue,nishihara,bhattacharyya}
The magnetism is unusual in that it has a high Curie temperature of
$T_C \sim$ 54 K
relative to the moment size, reported as $m$ = 0.06 $\mu_B$ -- 0.08 $\mu_B$
per Ni. \cite{lemaire,buschow}
If one scales the Curie temperature of elemental Ni ($T_C$=627 K),
which is also high relative to its moment size, by the square of the
moment as usual, one would infer an expected Curie temperature
of only $\sim$ 6--7 K for Y$_2$Ni$_7$.
The system is also unusual in that in spite of the high $T_C$,
it is very sensitive to alloying, both by H incorporation,
\cite{buschow} and by metal alloying.
\cite{levitin,ballou}
Here we investigate the magnetic and electronic properties of Y$_2$Ni$_7$
in relation to Ni metal as well as the weak itinerant ferromagnet Ni$_3$Al
and the related compound Ni$_3$Ga. We find behavior reminiscent of
Ni$_3$Al.

\section{Methods and Structure}

\begin{figure}
\includegraphics[width=\columnwidth,angle=0]{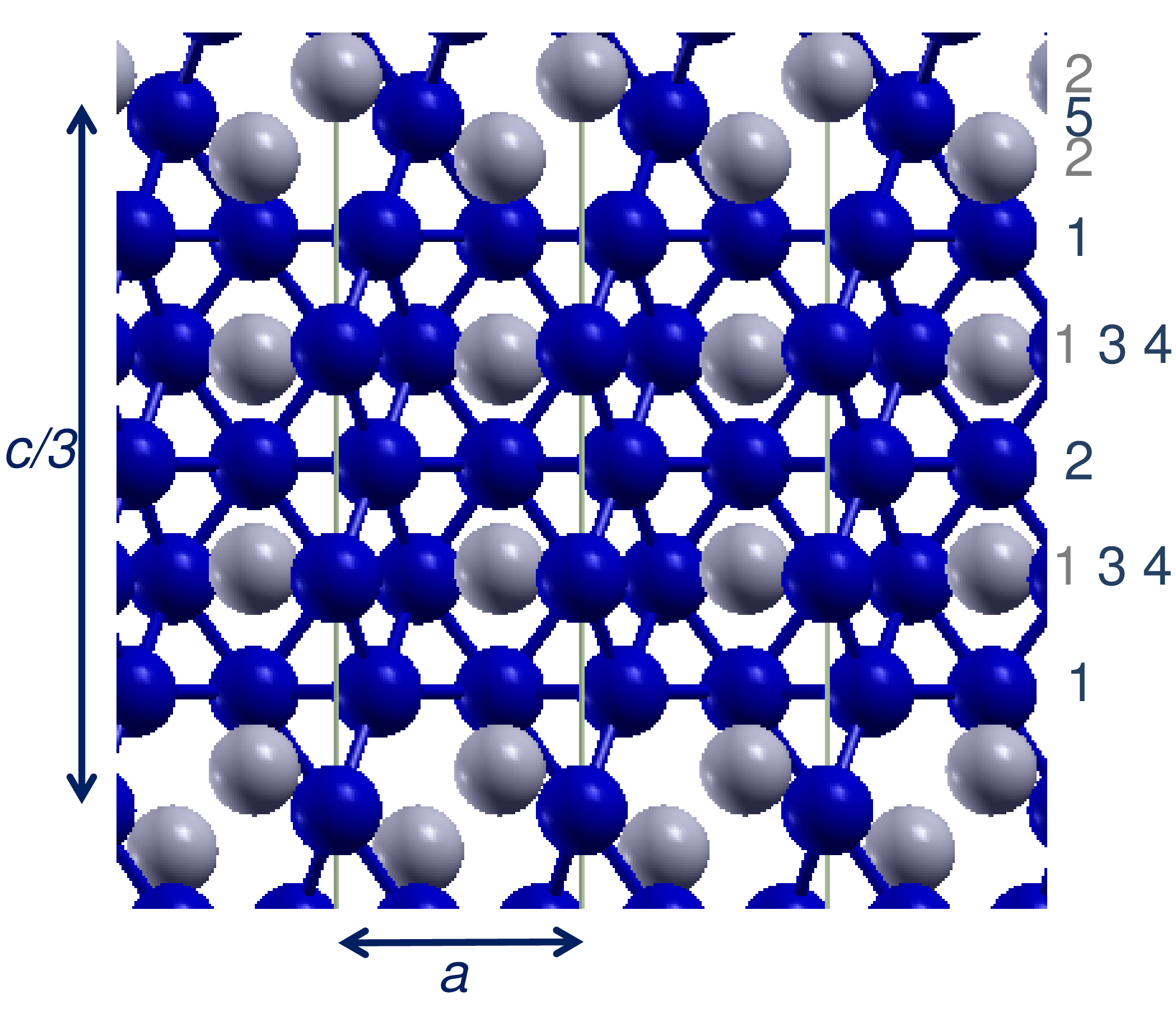}
\caption{Crystal structure of Y$_2$Ni$_7$ showing the layering of the
metal atoms. Y is shown as light grey, with Ni as dark blue. The numbers
denote the atom number as in Table \ref{tab-struct}.}
\label{fig-struct}
\end{figure}

The present density functional calculations were done using the 
generalized gradient approximation of Perdew, Burke and Ernzerhof
\cite{pbe} and the linearized augmented planewave (LAPW)
method \cite{singh-book} as implemented in the WIEN2k code.
\cite{wien2k}
We also performed calculations using the local spin density approximation
(LSDA) and we also tested the effects of spin-orbit coupling.
We used well converged basis sets, with a planewave cutoff,
$K_{max}$ determined by
$R_{min}K_{max}$=9, where $R_{min}$ is the smallest sphere radius, here
2.25 Bohr for both Y and Ni. The calculations were done using the
experimental lattice parameters, \cite{colinet}, $a$=4.947 \AA,
$c$=36.25 \AA.
The internal coordinates were determined by total energy minimization.
For this purpose we started with the structure of the prototype (Gd$_2$Co$_7$).
The resulting structure is given in Table \ref{tab-struct} and depicted in
Fig. \ref{fig-struct}.
Plasma frequencies were determined using the optical package of the
WIEN2k code, which uses integration of the squared band
velocity on the Fermi surface for this purpose. In this code the
velocities come from calculations of the dipole (momentum) operator.

\section{Results and Discussion}

Y$_2$Ni$_7$, which is $\sim$78\% Ni, has metal atoms in distorted 12-fold cages.
As mentioned,
Ni$_3$Al, which contains 75\% Ni, also
with a trivalent element
has suppressed ferromagnetism relative to Ni
and is near an interesting quantum critical point.
\cite{niklowitz,smith,hayden}
Ni$_3$Al has $T_C$=41.5 K, $M$=0.08 $\mu_B$/Ni, similar to Y$_2$Ni$_7$.
One signature of physical importance of fluctuations associated with the
quantum critical point in Ni$_3$Al is an overestimate of the ordered moment
in standard density functional calculations. In the case of Ni$_3$Al, the
calculated spin moment in the local density approximation is 
$M_{LDA}$=0.24 $\mu_B$/Ni. \cite{aguayo}

\begin{figure}
\includegraphics[width=\columnwidth,angle=0]{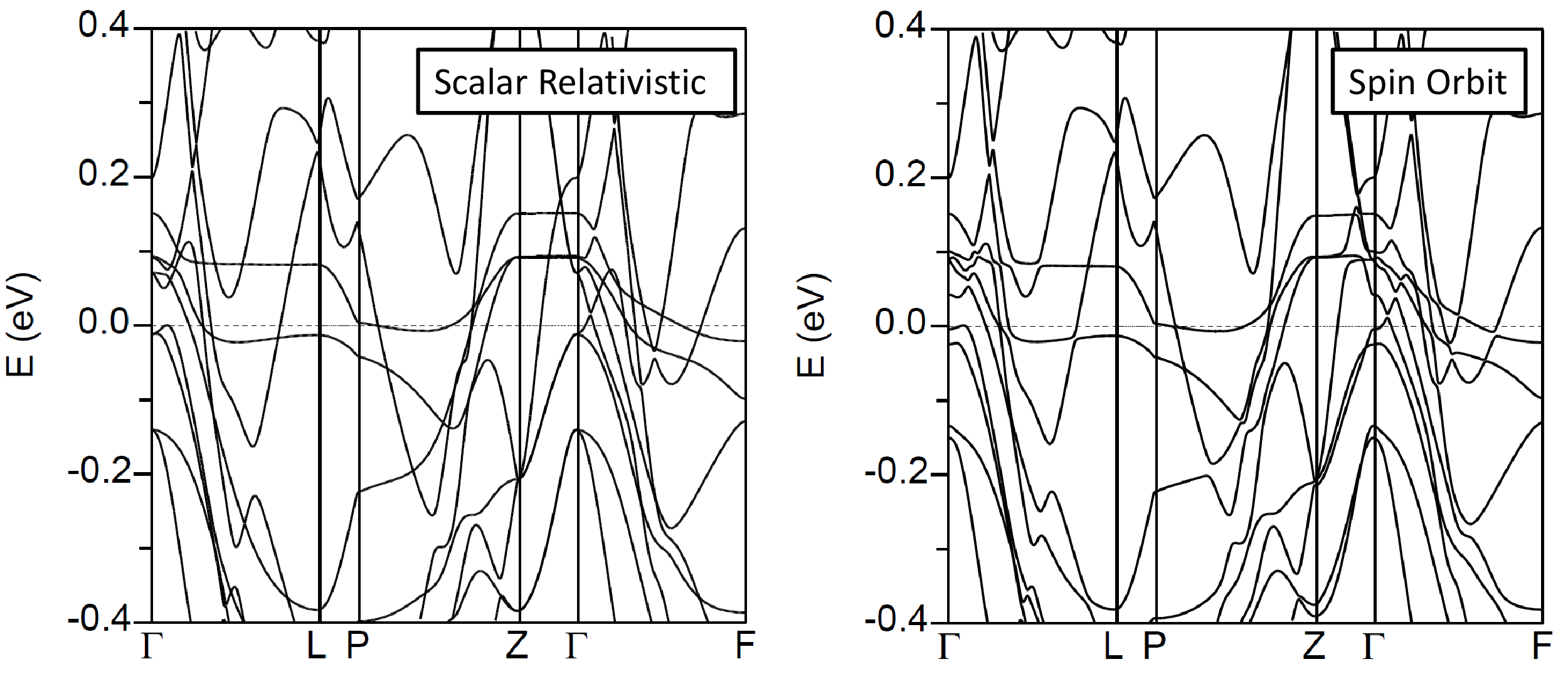}
\caption{Band structure of non-spin-polarized
Y$_2$Ni$_7$ as obtained with the PBE GGA
in a scalar relativistic approximation (left) and with spin-orbit
coupling (right).}
\label{bands}
\end{figure}

\begin{figure}
\includegraphics[width=\columnwidth,angle=0]{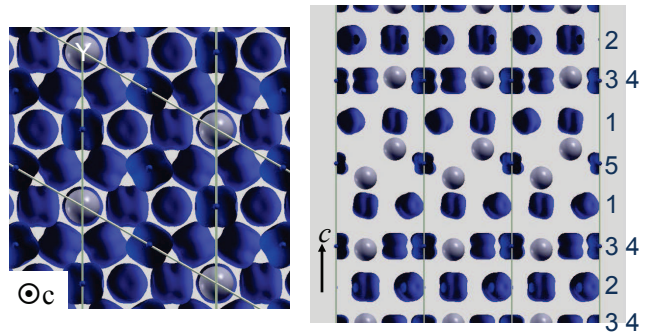}
\caption{Spin density plot showing the orbital character of the magnetization.
The left panel shows a view along the $c$-axis while the right panel is
perpendicular to the $c$-axis. The labels indicate the Ni atoms each layer
along $c$, as in Fig. \ref{fig-struct}.
Note the pronounced anisotropy around the Ni atoms.}
\label{fig-spindens}
\end{figure}

For Y$_2$Ni$_7$ we obtain a spin magnetization of 1.29 $\mu_B$ per
formula unit (f.u.) with the PBE GGA, with a magnetic energy of
0.04 eV/f.u. On a per Ni basis, this is 0.18 $\mu_B$ and
5.7 meV $\sim$ 70 K, i.e. only slightly higher than $T_C$=54 K.
This is indicative of being in the strongly itinerant (Stoner) limit.
It is notable that the moment size is $\sim$ 0.1 $\mu_B$ higher than
the reported experimental value.
The spin density is illustrated in Fig. \ref{fig-spindens}.
The moments on the different Ni sites, as determined by the spin magnetization
in the corresponding LAPW spheres,
vary considerably as seen
in Table \ref{tab-struct}. There is a small back polarization
in the interstitial and around the Y atoms, as is commonly 
found in 3$d$ transition metal ferromagnets.
This interstitial spin density is derived from extended orbitals, such as
the metal $s$ states.
The two Ni sites with low numbers of Ni neighbors (Ni1 and Ni5) have
the lowest moments, with Ni5 by far the lowest.
However, there is not a clear correlation between
coordination and the Ni moment for the other sites.

Calculations for fcc Ni done in the same way yield an ordered moment
of 0.635 $\mu_B$ and a magnetic energy of 0.062 eV/atom $\sim$740 K
(c.f. $T_C$=627 K).
Spin orbit coupling
has only a very small effect, reducing the spin moment by less
than 0.01 $\mu_B$/f.u. and producing orbital moments of 0.007 $\mu_B$
-- 0.025 $\mu_B$ per Ni, depending on the specific site, always aligned
with the spin moment, as expected from the third Hund's rule.
The effect of spin orbit coupling
on the electronic structure is also very small.
This is seen in the band structure near the Fermi energy, $E_F$, which
is given in Fig. \ref{bands} in a scalar relativistic approximation and
with spin orbit coupling.
In the following, we give scalar relativistic results.

With the LSDA we still obtain an
overestimation of the ordered moment, although somewhat smaller, specifically,
$M_{LSDA}$=1.17 $\mu_B$/f.u. and $\delta E_{LSDA}$=0.027 eV/f.u. or
$\sim$47 K on a per Ni basis, i.e. slightly less than $T_C$. This
similarity of the magnetic energy to the experimental $T_C$ is similar
to what was found for Ni$_3$Al, \cite{aguayo}
suggesting a similarity of Y$_2$Ni$_7$ and Ni$_3$Al.
However, it should be noted that the more complicated
non-cubic structure of Y$_2$Ni$_7$ may lead to larger experimental
uncertainty in the determination of the moment size due to magnetocrystalline
anisotropy and more possibilities for intrinsic defects.

\begin{table}
\caption{Calculated internal structural parameters
and atomic moments with the PBE GGA of Y$_2$Ni$_7$ at
the experimental lattice parameters of $a$=4.947 \AA, $c$=36.25 \AA,
space group 166, R$\bar{3}$m with hexagonal coordinates. The nearest
12 neighbors are given as ``coord.".}
\begin{tabular}{lccccc}
\hline
Atom & $x$ & $y$ & $z$ & coord. & m($\mu_B$) \\
\hline
Y1  6$c$ & 0 & 0 & 0.0504 & 12 Ni & -0.03 \\
Y2  6$c$ & 0 & 0 & 0.1472 & 12 Ni & -0.02 \\
Ni1 18$h$ & 0.1671 & 0.8329 & 0.4430 & 7 Ni, 5 Y & 0.17 \\
Ni2 9$e$ & 1/6 & 1/3 & 1/3 & 8 Ni, 4 Y & 0.33 \\
Ni3 6$c$ & 0 & 0 & 0.2782 & 9 Ni, 3 Y & 0.22 \\
Ni4 6$c$ & 0 & 0 & 0.3883 & 9 Ni, 3 Y & 0.23 \\
Ni5 3$b$ & 0 & 0 & 1/2 & 6 Ni, 6 Y & 0.10 \\
\hline
\end{tabular}
\label{tab-struct}
\end{table}

\begin{figure}
\includegraphics[width=\columnwidth,angle=0]{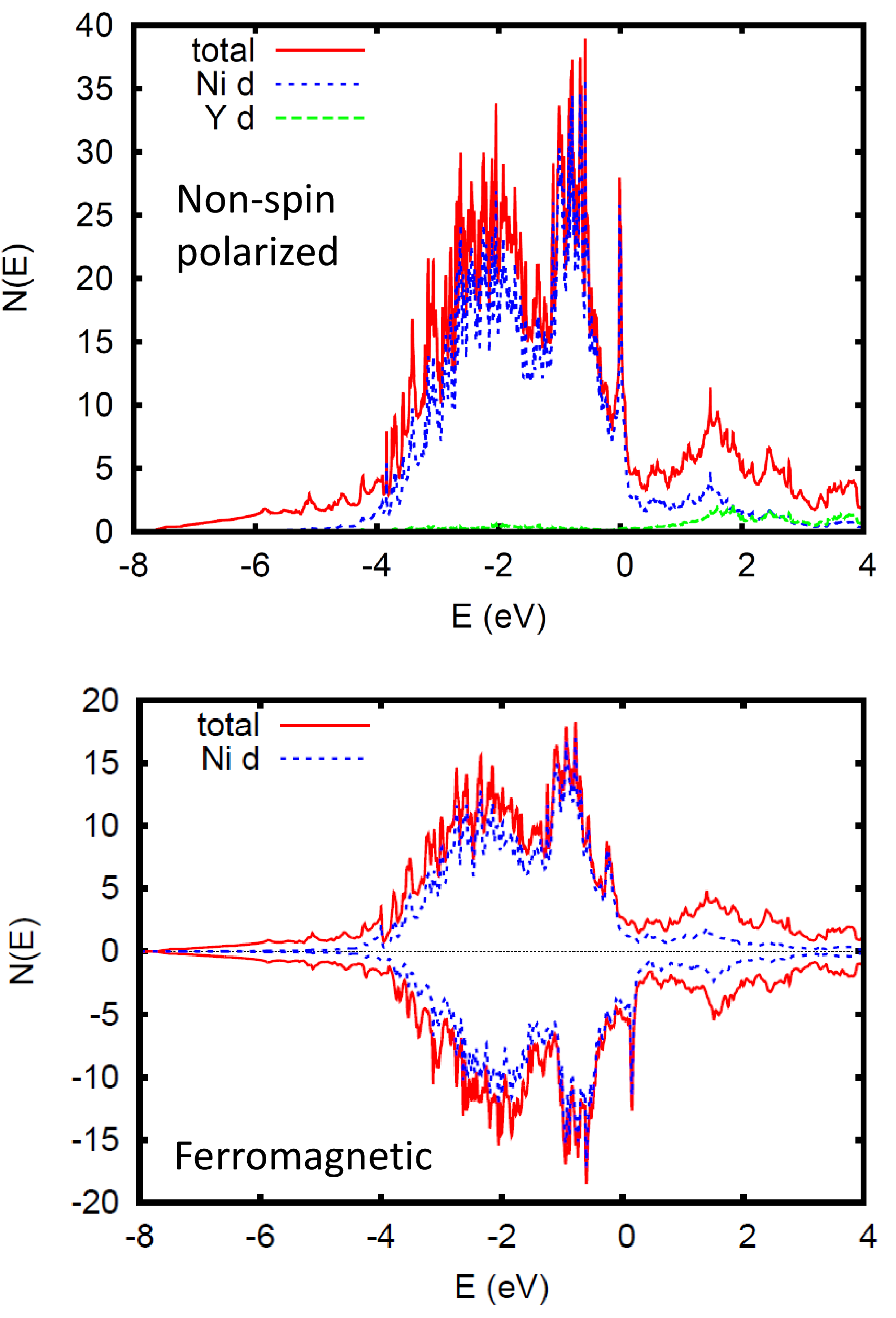}
\caption{Electronic density of states of Y$_2$Ni$_7$ as obtained with
the PBE GGA for non-spin-polarized (top) and ferromagnetic (bottom) states.}
\label{fig-dos}
\end{figure}

The calculated electronic density of states and projections
of Ni $d$ and Y $d$ character are given in Fig. \ref{fig-dos},
both for the non-spin-polarized and the ferromagnetic cases.
As shown, there is a sharp peak in the density of states almost
exactly at $E_F$. The high value leads to a Stoner instability
and ferromagnetism, with an exchange splitting of the Ni $d$ bands, 
although not in a perfect rigid band fashion (note the change in the
shape of the exchange split peak between majority and minority spin).
Also, the Y $d$ states are above $E_F$ in this compound. In contrast,
Y metal is a transition element with $\sim$ 2 $d$ electrons. This
implies a charge transfer from the Y $d$ states to the Ni host matrix.
This is not surprising in view of the electropositive nature of Y
relative to Ni.
Based on integration of the Ni $d$ density of states in comparison with
fcc Ni, most of this charge transfer is to the $s$ electrons. This inference
is based on the fact that the $d$ electron count of Ni does not show
an increase commensurate with the additional charge.

Within Stoner theory \cite{stoner,gunnarsson}
the susceptibility of a metal is given by
a random phase approximation (RPA) formula,
$\chi=\chi_0/[1-N(E_F)I]$, where $\chi_0$ is the bare
Pauli susceptibility, $\chi_0=\mu_B^2 N(E_F)$, with appropriate units.
This formula is exact at the level of band structure calculations,
but neglects the effects of spin-fluctuations which can renormalize
the spin susceptibility (see Ref. \onlinecite{larson} for a detailed
discussion applied to Pd, which is a high susceptibility paramagnetic metal).
The Stoner theory
itinerant ferromagnetic instability occurs when $N(E_F)=I^{-1}$, which is the
point where the RPA enhancement factor $1/[1-N(E_F)I]$ diverges.

In contrast to standard local moment magnetic materials, in the itinerant
limit a metal with properties determined
by the non-spin-polarized electronic structure occurs above $T_C$. There is
typically a second order or near second order phase transition at $T_C$.
In Y$_2$Ni$_7$ this is a rather interesting metallic state in part
because of the high density of states.

\begin{figure}
\includegraphics[width=\columnwidth,angle=0]{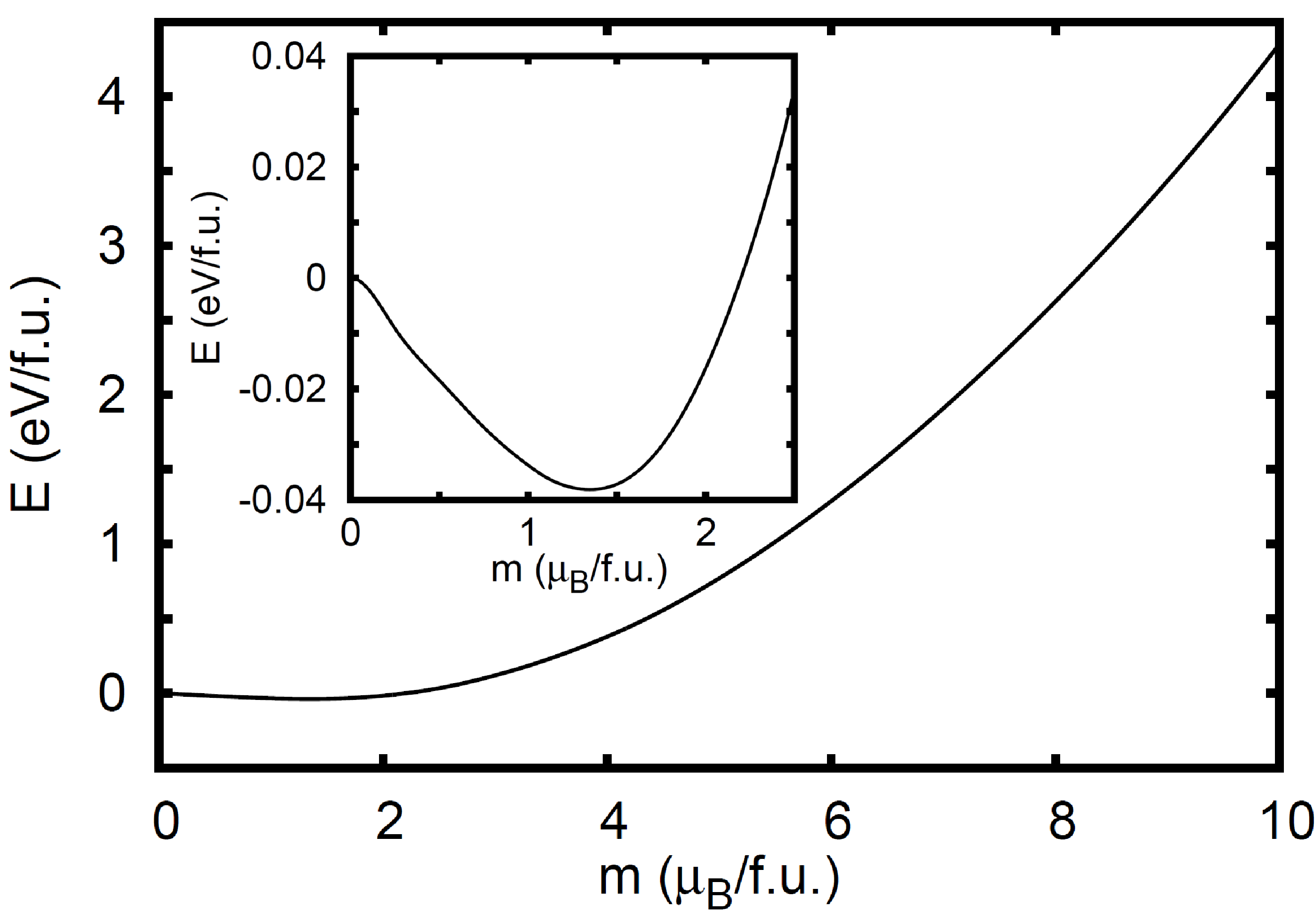}
\caption{Fixed spin moment energy as a function of constrained magnetization.
The inset shows an expanded scale for low magnetizations.}
\label{fig-fsm}
\end{figure}

We obtain $N(E_F)$=24.0 eV$^{-1}$ per f.u. on a both spins
basis. This well above the criterion for Stoner magnetism,
\cite{stoner,gunnarsson,moriya}
$N>I^{-1}$ where $I$ is 0.7-0.9 eV for heavy
3$d$ transition elements \cite{janak} and $N$ is $N(E_F)$
expressed on a per atom per spin basis (1.71 eV$^{-1}$ in the present case).
The fixed spin moment energy (Fig. \ref{fig-fsm})
as a function of magnetization therefore drops
quickly from zero but then rises because of the narrowness of the peak.
The curve is featureless and smooth except for the minimum corresponding
to the ferromagnetic ground state. There is no sign of any metamagnetic
state at higher magnetization.

The peak at $E_F$ is derived from Ni $d$ orbitals on the Ni2, Ni3 and
Ni4 sites, with a somewhat smaller contribution from the Ni1 site and
practically no contribution from the Ni5 site. This mirrors the
distribution of the moments in the ferromagnetic state. For an atom with
trigonal site symmetry (i.e. axial with a three fold axis), as are the
Ni3, Ni4 and Ni5 atoms, the $d$ electron crystal field levels are a singly
degenerate $a_g$ ($z^2$, with $z$ along the axis), and two double
degenerate $e_g$ levels ($x^2-y^2$+$xy$ and $xz$+$yz$).
The Ni3 and Ni4 contributions to the peak at $E_F$ are from the
$e_g$ ($xz$+$yz$) orbital. The Ni1 and Ni2 contributions are from the same
two $d$ orbitals in this reference frame, but not in equal proportion
reflecting the lower site symmetry. The orbital character is reflected in the
spin density (Fig. \ref{fig-spindens}), where the lobes corresponding to these
orbitals are seen around the Ni sites. As seen in the figure these orbitals
are not oriented favorable for bonding interactions, which is a fact
consistent with the narrowness of the peak.

The magnetic
behavior found in the calculations
is consistent with what is expected from extended Stoner theory
based on the sharply peaked density of states.
Within extended Stoner theory, \cite{andersen-es,krasko-es}
which involves a rigid band approximation,
one exchange splits the density of states to obtain magnetization.
The stationary solutions are points for which
$\overline{N(m)}=I^{-1}$, where $\overline{N(m)}$ is the average
density of states between the position of the Fermi level for minority
spin and that for majority spin to obtain the magnetization, $m$
using the non-spin-polarized density of states.

In other words $\overline{N(m)}$ is the magnetization divided
by the energy shift between majority and minority spin densities of 
states needed to produce this magnetization.
Thus a sharp peak with little area underneath it will yield a strong
initial ferromagnetic instability in a fixed spin moment plot, but will 
not lead to a high magnetization. This is the origin of the high
$T_C$ with low moment.
In other words a narrow peak with a
very high $N(E_F)$ leads to a strong initial instability, i.e.
a large negative $\chi=\chi_0/[1-N(E_F)I]$. However if the weight under
the peak is small, the net magnetization will be low because
$\overline{N(m)}$ will not be large for finite $m$ since it is an
integral.
Note also that the extended Stoner formula uses the connection between
the moment size and the exchange splitting through the fact that the moment
is an integral of the density of states over the energy range coming from
the exchange splitting. This connects the moment, the exchange splitting
and the magnetic energy.

The density of states at the Fermi level is reduced to
$N(E_F)$=9.1 eV$^{-1}$ per f.u., mainly from the minority spin in the
ferromagnetic ground state.
The majority and minority spin contributions are
$N_\uparrow(E_F)$= 2.8 eV$^{-1}$ and $N_\downarrow(E_F)$=5.3 eV$^{-1}$,
respectively.
This corresponds to a bare specific heat coefficient,
$\gamma_{bare}$=18.9 mJ/(mol K$^2$) on a per f.u. basis.
It would be interesting to compare with experiment to determine the
specific heat enhancement, $\gamma/\gamma_{bare}$,
which if large might be an indicator of
quantum spin fluctuations.

We now return to the non-spin-polarized case, which should correspond to the
electronic structure above $T_C$ in this itinerant material.
The band structure (Fig. \ref{bands}) shows heavy bands close to $E_F$,
specifically the band just below $E_F$ along a large part of the $\Gamma$--$L$
line and the band at $E_F$ along $P$--$Z$. In addition there are several
much more dispersive bands crossing $E_F$. The consequence is that although
the value of $N(E_F)$ is high, the material can have a reasonable
conductivity.
This multi-sheet Fermi surface characteristic also occurs in Ni$_3$Ga
and Ni$_3$Al. \cite{hayden,aguayo}
In general the conductivity can be written as
$\sigma\propto\omega_p^2 \tau$, where $\tau$ is an effective inverse
scattering rate and $\omega_p$ is the plasma frequency.
We obtain plasma energies $\hbar\omega_p$ of
$\Omega_{p,a}$=2.3 eV and $\Omega_{p,c}$=2.0 eV, for the basal plane
and $c$-axis directions, respectively. The implied conductivity
anisotropy is modest, $\sigma_a/\sigma_c\sim$1.3.
The anisotropy increases in the low temperature ferromagnetic state,
for which we obtain $\sigma_a/\sigma_c\sim$1.8.
Also, even though the magnetization is small we obtain a significant
transport spin polarization, defined by 
$P=(\sigma_\uparrow-\sigma_\downarrow)/(\sigma_\uparrow+\sigma_\downarrow)$.
We obtain $P$=0.09 both in the basal plane and $c$-axis directions.

As mentioned, Y$_2$N$_7$ has a remarkably high ordering temperature in
relation to its moment. Most magnetic materials are described in terms
of local moments and their interactions through the interatomic
exhange couplings, $J_{i,j}$. These describe spin wave dispersions, which
are transverse in character. Longitudinal degrees of freedom, which
correspond to changes in the local moment size are hard and
not involved in the phase transition. This leads to a simplification in
which one can treat the phase transition using the effective spin
Hamilitonian and the effect of increasing temperature as the excitation
of spin-waves.
This local moment case has been well described theoretically and
numerical simulations of the temperature dependent magnetic properties
and phase transitions are practical even
for complex systems. \cite{liechtenstein,sato}
Importantly, the paramagnetic state above the ordering
temperature is a disordered local moment state in which the local moment
directions may be regarded as fluctuating in time but retaining their size.
This means that the atomic Hund's rule energy associated with the moment
formation (i.e. the longitudinal degree of freedom) is not involved in the
phase transition as this
contribution to the magnetic energy
is present in both the ordered and paramagnetic phases.

The itinerant limit has excitation of both transverse and longitudinal
degrees of freedom with temperature
through coupling to the electronic system and is not
describable by an effective spin Hamiltonian.
This is more difficult to treat theoretically as it involves
coupling to the electrons without separation of electronic and
magnetic degrees of freedom.
Y$_2$Ni$_7$ and Ni$_3$Al
are both apparently close to this limit, and based on comparison of
density functional results with experiment both also have renormalizations
of their ground states due to quantum fluctuations. Itinerant magnets
also have magnetic contributions to the energy in the paramagnetic
state, \cite{staunton} above but close to the ordering temperature.
However, these are reduced as the moment size is reduced, and in the
itinerant limit become negligible. This means that in the itinerant limit
all of the magnetic energy including the onsite Hund's energy is available
to drive the ordering, providing an explanation for the high ordering 
temperatures relative to the moments in these itinerant materials.
In other words, in itinerant magnets disordering implies destroying the
moments, which has an energy cost from the on-site Hund's energy and
this leads to high ordering temperatures.
We note that the Stoner model of itinerant magnetism has a parallel in
the Slater model of itinerant antiferromagnetism and that high ordering
temperatures in certain antiferromagnets have been discussed in a way similar
to the above. \cite{rodriguez,shi,singh-SrRu2O6}
In neither case (itinerant or local moment magnets) can energy
differences by themselves be simply interpreted as the ordering
temperature.

\section{Summary and Conclusions}

We report density functional calculations of the electronic structure
and magnetic properties of Y$_2$Ni$_7$.
Y$_2$Ni$_7$ shows similarity to the weak itinerant ferromagnet Ni$_3$Al,
which has a modest renormalization of the magnetism due to nearness
to a quantum critical point.
The overestimation of the magnetization
relative to experiment is a bit smaller in Y$_2$Ni$_7$ implying
perhaps somewhat weaker
fluctuation effects. This is in contrast to Ni$_3$Ga, where magnetic
ordering is apparently completely suppressed by quantum spin fluctuations.
The weak itinerant ferromagnetism in Y$_2$Ni$_7$ arises from a Stoner
instability of a rather interesting metallic state. This state features
a very narrow density of states peak at $E_F$, a mixture of dispersive
and flat bands crossing $E_F$ and modest but
non-negligible anisotropy of the plasma frequency.
Thus Y$_2$Ni$_7$ is a uniaxial analogue of cubic Ni$_3$Al.
It will be of interest to study the low temperature properties of 
Y$_2$Ni$_7$, its magnetic fluctuations and the pressure dependence
of the magnetic, thermodynamic and
transport properties in comparison with Ni$_3$Al.

\acknowledgments

A portion of this work was performed at Oak Ridge National Laboratory
with support from the
Department of Energy, Basic Energy Sciences,
Materials Sciences and Engineering Division.

\bibliography{Y2Ni7}

\end{document}